\documentclass[12pt]{article}
\usepackage{geometry}
\usepackage{amssymb}
\usepackage{amsmath}
\usepackage{bbm}
\usepackage{MnSymbol}
\usepackage{hyperref}
\usepackage{braket}
\usepackage{cite}
\usepackage{subcaption,tikz}
\usetikzlibrary{arrows,decorations.markings,decorations.pathreplacing}

\newcommand{\be}{\begin{equation}}
\newcommand{\ee}{\end{equation}}
\newcommand{\bea}{\begin{eqnarray}}
\newcommand{\eea}{\end{eqnarray}}

\newcommand{\om}{\omega}

\newcommand{\io}{\iota}
\newcommand{\mcL}{\mathcal{L}}

\newcommand{\lrr}{\longrightarrow}

\newcommand{\Z}{\mathbb{Z}}

\newcommand{\Hom}{\operatorname{Hom}}

\newcommand{\Rep}{\operatorname{Rep}}
\newcommand{\Ve}{\operatorname{Vec}}

\tikzset{
	partial ellipse/.style args={#1:#2:#3}{
		insert path={+ (#1:#3) arc (#1:#2:#3)}
	}
}

\numberwithin{equation}{section}

\begin{document}
	
\begin{flushright}
	MI-HET-796
\end{flushright}
	
\begin{center}

{\large\bf Perspectives on Anomaly Resolution}

	\vspace*{0.2in}
	
	Thomas Vandermeulen
	
	\vspace*{0.2in}
	
		{\begin{tabular}{l}
				George P.~and Cynthia W.~Mitchell Institute\\
				for Fundamental Physics and Astronomy\\
				Texas A\&M University\\
				College Station, TX 77843 \end{tabular}}
			
	\vspace*{0.2in}
	
	{\tt tvand@tamu.edu}
	
\end{center}
\pagenumbering{gobble}

We examine in detail the process of resolving 't Hooft anomalies by extending the symmetry of a theory.  Specifically, we interpret the ingredients of existing prescriptions for anomaly resolution as the addition of topological operators with designated mixed anomalies, which can be interpreted as coupling our original field theory to a topological one.  We show that, upon gauging, the presence of such mixed anomalies leads to a modified version of the original symmetry which now acts on the newly introduced operators, allowing for an overall anomaly-free action.  We also show that the original, anomalous symmetry remains present in the theory.  This analysis is applied to anomaly-resolving extensions by both ordinary and higher-form symmetries, leading to related but qualitatively distinct stories.

\newpage
\tableofcontents

\newpage

\section{Introduction and Background}
\label{sec:intro}
\pagenumbering{arabic}

One sometimes finds obstructions when attempting to promote a global symmetry $G$ to a gauge symmetry.  In the case that $G$ is a bosonic symmetry given by a finite discrete group, which will be the main case of interest in this paper, such anomalies are known as 't Hooft anomalies and classified by the group cohomology $H^{d+1}(G,U(1))$ where $d$ is the spacetime dimension of the theory.  Specifically, such an anomaly manifests as an obstruction to coupling the theory to a background $G$ field in a gauge-invariant manner.  In modern terms, we would realize such a gauge field as a network of topological defects living on the worldsheet of the theory.  The anomaly then manifests as phase inconsistencies which arise when resolving higher junctions.\\

The most obvious way to avoid a non-trivial anomaly $\omega\in H^{d+1}(G,U(1))$ of $G$ is to pick a non-anomalous subgroup $G_0$ of $G$ and gauge $G_0$ instead of $G$.  More specifically, because $G_0$ is a subgroup we have an inclusion homomorphism $\io:G_0\to G$.  For $G_0$ to be non-anomalous, the pullback $\io^*\omega$ should be in the trivial class of $H^{d+1}(G_0,U(1))$.

However, this is not the only option.  \cite{WWW} provides an alternative method for resolving anomalies by extending, rather than restricting, the symmetry $G$.  This approach involves identifying a trivially-acting symmetry $K$ of the theory such that the total symmetry is now an extension of $G$ by $K$, rather than $G$ by itself.  That is, there should exist a short exact sequence
\be
1 \lrr K \xrightarrow{\hspace{0.2cm}\io\hspace{0.2cm}} \Gamma \xrightarrow{\hspace{0.2cm}\pi\hspace{0.2cm}} G \lrr 1
\ee
such that, if $\pi$ is the homomorphism $\Gamma\to G$, the pullback $\pi^*\omega$ is trivial in $H^{d+1}(\Gamma,U(1))$.  Again we trivialize the anomaly via pullback, but the interpretation is now very different -- we have added to our symmetry, instead of ignoring part of it.\\

The purpose of this work is to examine the above construction in detail, and give an explanation of how the presence of the trivially-acting symmetry $K$ facilitates the construction of a non-anomalous $G$ symmetry from an anomalous one.  The broad explanation upon which we will settle relies on the perspective advocated in \cite{TopOps} that a trivially-acting 0-form symmetry should be regarded as a mix of a 0-form and a 1-form symmetry.  This opens up two possibilities for $G$ to interact with $K$ -- it can mix non-trivially (in the sense of a group extension) with the 0-form part of $K$, and it can have a mixed anomaly with the 1-form part of $K$.  This interplay will allow us to, in some sense, separate $G$ from its anomaly via its interactions with the codimension 2 defects that control the 1-form part of $K$.

While the above situation is generic in spacetime dimension $d>1$, it will nonetheless be helpful to restrict to specific low dimensions in looking at examples.  Section~\ref{sec:2d} will begin in two dimensions, in which this story exhibits some rather special features.  Because this is the unique dimension for which the 1-form part of $K$ constitutes a $(d-1)$-form symmetry, we will find that the anomaly resolution procedure involves the consideration of multiple copies of the theory in question.  Additionally, in 2d conformal field theory (CFT), 't Hooft anomalies manifest as obstructions to modular invariance of the gauged theory.  This will make it quite straightforward to see anomalies resolved in specific examples.  Section~\ref{sec:3d} runs through the story in three dimensions, which is almost entirely generic, with the sole standout feature that the gauged theory contains only one-form symmetries (in all other dimensions the result will be a mixed symmetry).  

Finally, section~\ref{sec:1d} pivots to the special case of one dimension, which will have a unique variant of the story, as there are no 1-form symmetries to utilize.  We will find that the anomaly resolution in this case can be understood in terms of $-1$-form symmetries, as explored in \cite{lfs}.  The physical picture developed in this section will apply generally to the procedure given in \cite{Kobayashi_2019} in which one resolves an anomalous 0-form symmetry through extension by a $(d-1)$-form symmetry.  We will also see how a similar procedure can resolve a 1-form symmetry in three dimensions.

\subsection{Notation and Conventions}
\label{sec:conventions}

We will frequently regard symmetries in terms of the topological operators that implement them.  As we will be concerned exclusively with topological operators, when we refer to e.g.~`lines' in a theory we will mean topological defect lines, and similarly for operators of other dimension.  Generally we will leave the topological property of operators implied in lieu of constantly specifying it.\\

Any generic groups appearing in this paper are assumed to be discrete. Specific non-discrete groups may still appear (for instance, we will often use group cohomology valued in $U(1)=S^1$).  Group cohomology is taken, as usual, to be the singular cohomology of the classifying space, i.e.~$H^p_{\text{group}}(G,K)=H^p_{\text{sing.}}(BG,K).$\\

A $p$-form symmetry given by a group or fusion category $G$ will be written as $G_{[p]}$.  A dot will be used to indicate a general symmetry extension.  That is, for groups $G$ and $K$, $K.G$ means an extension (not necessarily split, not necessarily central) of $G$ by $K$.  For more general symmetries the same notation $K_{[q]}.G_{[p]}$ indicates that we have a mix of a $q$-form symmetry described by $K$ (which may not necessarily be group-like) and a $p$-form symmetry described by $G$.  A `non-trivial extension,' in this context, would obstruct $G_{[p]}$ from being a subsymmetry of the theory.  This obstruction can be alternatively viewed as a generalization of the extension class, or a form of anomaly.\footnote{Adopting this language, an `extension class' for $K_{[q]}.G_{[p]}$ would be an obstruction to gauging $G_{[p]}$ but not $K_{[q]}$ or the total symmetry, while a `mixed anomaly' is an obstruction to gauging the total symmetry but not $K_{[q]}$ or $G_{[p]}$ individually.}

\subsection{Symmetry Extension as Coupling to TQFT}
\label{sec:tqft}

The notion of altering the symmetries of a theory by introducing new, trivially-acting symmetries may seem vague.  At the least this process begs questions such as `are we now discussing an entirely different theory' and `should we regard theories as automatically including all possible trivially-acting symmetries?'  To be able to address such concerns, we would like to have a physical picture of such a process.  Fortunately, there is a known trick to accomplish just this -- if one begins with a quantum field theory (QFT) and couples it to a topological quantum field theory (TQFT), one can obtain a theory with the same local dynamics but an altered extended spectrum \cite{KapustinSeiberg,Seiberg_2010,GaiottoKulp,Unsal,SP1,SP2,SP3}.  We would like to interpret the symmetry extensions used here as such a process.\\

Perhaps the best-known example of this procedure appears in orbifolds in 2d CFT.  With an orbifold group $G$, the resulting torus partition function nominally takes the form
\be
\label{orbpf}
\frac{1}{|G|}\sum_{g_1,g_2\in G}Z_{g_1,g_2}.
\ee
There is, however, a well-known freedom possessed by the phases appearing in this sum.  Given an element $\om$ representing a class in $H^2(G,U(1))$, one can modify (\ref{orbpf}) to
\be
\label{orbpf_dt}
\frac{1}{|G|}\sum_{g_1,g_2\in G}Z_{g_1,g_2}\frac{\om(g_1,g_2)}{\om(g_2,g_1)}
\ee
while maintaining modular invariance at all genera.  Thus $H^2(G,U(1))$, known as discrete torsion in the $G$ symmetry, classifies different consistent $G$ gaugings of the theory \cite{VafaDT}.

Now recall that the ratio of cocycles appearing in (\ref{orbpf_dt}) has an alternative interpretation.  It is the weight assigned to the $g_1,g_2$ sector of a $G$-symmetric symmetry-protected topological (SPT) phase, which is an invertible QFT (i.e.~its Hilbert space on any manifold is one-dimensional) carrying $G$ symmetry.  Such theories are classified in two dimensions by $H^2(G,U(1))$ \cite{ChenGuLiuWen}.  The upshot of this observation is that, assuming we have a theory whose genus one partition function on a torus wrapped by line operators implementing $g_1$ and $g_2$ is given by
\be
S_{g_1,g_2}=\frac{\om(g_1,g_2)}{\om(g_2,g_1)},
\ee
we can rewrite (\ref{orbpf_dt}) as
\be
\label{spt_stack}
\frac{1}{|G|}\sum_{g_1,g_2\in G}Z_{g_1,g_2}S_{g_1,g_2}
\ee
and interpret the $G$ orbifold with discrete torsion as gauging the diagonal $G$ symmetry of the direct product of our original theory with the $G$-SPT described by the cocycle $\om$.  This is often referred to as `stacking' with an SPT -- see for example \cite{GaiottoKulp,HsinLam,LOZ}.\\

Now let us modify the above construction.  While taking the diagonal $G$ symmetry above was a convenient way to reproduce discrete torsion phases, it certainly wasn't necessary.  In fact, let us begin with a local theory whose symmetries we will ignore and couple it to a $G$-SPT.  For simplicity we will take $G$ abelian and $\om$ to be in the trivial class in $H^2(G,U(1))$.  In such a situation (\ref{spt_stack}), which will now be the orbifold of only the SPT, becomes
\be
\frac{1}{|G|}\sum_{g_1,g_2\in G}ZS_{g_1,g_2} = \frac{1}{|G|}\sum_{g_1,g_2\in G}Z = |G|Z
\ee
from which we infer that the resulting theory is the direct sum of $|G|$ copies of our original, local CFT.  As gauging a SPT produces a Dijkgraaf-Witten (DW) model \cite{DW}, we can alternatively regard this direct sum of $|G|$ copies of a theory as coupling to a 2d DW model with symmetry $G$ and trivial cocycle $\om$, which is itself isomorphic to $|G|$ copies of a trivial theory (see e.g.~\cite[section 6.6]{UndoingDecomp}).\footnote{Note that if we took $G$ to be nonabelian we would instead find a copy of the theory for each conjugacy class in $G$.  The interpretation in terms of DW models becomes slightly more interesting, as the resulting theory is no longer a direct sum of trivial theories, but now can include non-trivial SPTs.  See \cite[section 4.4]{TopOps} for an example.}

This is exactly the result that we would have found by supplementing our local theory with a trivially-acting $G$ symmetry.  In general we could imagine beginning with a $G$-symmetric theory and coupling it to a $K$-SPT, with cohomological parameters such as the `extension class' $H^2(G,K)$ classifying the ways in which the $K$ symmetry operators could interact with the $G$ ones.  Such a picture should apply equally well to higher-form symmetries, so long as we choose a TQFT with suitable topological operators.  This is the sort of setup we should keep in the back of our mind when working with symmetry extensions.

\subsection{Extension Swapping}
\label{sec:extswap}

In order to examine anomaly resolution in greater detail, we will need to work with theories and their gaugings at the level of symmetries.  This section develops some basic technology for doing so.  It should be a familiar fact that when one gauges a discrete symmetry group $\Gamma$, the gauged theory possesses a dual `quantum' symmetry given by the fusion category $\Rep(\Gamma)$ formed by the irreducible representations of $\Gamma$ \cite{FRS,BhardwajTachikawa,Chang:2018iay}.  Many of the symmetries we will consider will come as group extensions, which is to say that $\Gamma=K.G$ fits into a short exact sequence
\be
\label{gamma_ext}
1 \lrr K \lrr \Gamma \lrr G \lrr 1.
\ee
When $\Gamma$ takes such a form, what can we say about $\Rep(\Gamma)$?  In order to answer this, let $T$ be a theory with symmetry $\Gamma$, which we assume to be non-anomalous.  Since $K$ is a normal subgroup of $\Gamma$, we can gauge $K$ alone.  What are the symmetries of the resulting theory $T/K$?  \cite{Tachikawa} provides the answer: $T/K$ has as its symmetry a direct product
\be
\label{gauge_sym1}
\Rep(K)\times G,
\ee
with a mixed anomaly determined by the extension class in (\ref{gamma_ext}).  This phenomenon of gauging interchanging extension classes and mixed anomalies is important, and will reappear frequently.  The upshot of the mixed anomaly is that we could gauge $\Rep(K)$ by itself (returning us to the original theory), we could gauge $G$ by itself (which we will do momentarily), but we generically cannot gauge the full $\Rep(K)\times G$ symmetry.

Now we gauge the $G$ symmetry of $T/K$, bringing us to $[T/K]/G$.  What is the symmetry of this theory?  In order to determine that, we can run in reverse the argument that produced (\ref{gauge_sym1}) from (\ref{gamma_ext}).  That is, the mixed anomaly between $G$ and $\Rep(K)$ becomes an extension class\footnote{Since we are allowing for non-group-like symmetries, the notion of `extension class' here may not correspond to a group cohomology class.}, and the symmetry of $[T/K]/G$ should be an extension $\Rep(G).\Rep(K)$.  The final connection to make in this analysis is that, because gaugings of a theory should compose, we in fact have $[T/K]/G=T/\Gamma$.  This allows us to identify the symmetry of $[T/K]/G$ as $\Rep(\Gamma)=\Rep(G).\Rep(K)$.

Summarizing, when a symmetry $\Gamma$ fits into a short exact sequence of the form (\ref{gamma_ext}), its dual symmetry $\Rep(\Gamma)$ fits into a dual short exact sequence
\be
\label{dual_ext}
1 \lrr \Rep(G) \lrr \Rep(\Gamma) \lrr \Rep(K) \lrr 1.
\ee
In discussion we will refer to this phenomenon as extension swapping, as $G$ and $K$ have reversed their positions between the extension (\ref{gamma_ext}) and the dual extension (\ref{dual_ext}).\\

This physical argument implies some purely mathematical facts.  For instance, let us take $\Gamma$ (and therefore $K$ and $G$ as well) to be finite abelian groups.  $\Rep(\Gamma)$ is then $\hat{\Gamma}$, the Pontryagin dual group to $\Gamma$.  $\Gamma$ is non-canonically isomorphic to $\hat{\Gamma}$, so in particular $\hat{\Gamma}$ is again a finite abelian group.  The dual extension (\ref{dual_ext}) takes the form
\be
1 \lrr \hat{G} \lrr \hat{\Gamma} \lrr \hat{K} \lrr 1.
\ee
The original extension (\ref{gamma_ext}) is classified by the group cohomology $H^2(G,K)$ (with trivial action on the coefficients).  In particular, since we assumed that $\Gamma$ is abelian, any particular $\Gamma$ comes with the specification of a symmetric class in $H^2(G,K)|_{\text{symm.}}$ -- that is, a $K$-valued cocycle $c(g_1,g_2)$ satisfying $c(g_1,g_2)=c(g_2,g_1)$.\footnote{A coboundary shift takes $c(g_1,g_2)$ to $c(g_1,g_2)+\lambda(g_1)+\lambda(g_2)-\lambda(g_1g_2)$ for some map $\lambda:G\to K$.  As $K$ and $G$ are both abelian, this shift is symmetric in $g_1$ and $g_2$, meaning that coboundary shifts preserve symmetric cocycles.  Thus we can sensibly speak of an entire cohomology class being symmetric or not.}  The dual group $\hat{\Gamma}$ similarly comes with a symmetric extension class $\bar{c}(\hat{k}_1,\hat{k}_2)\in H^2(\hat{K},\hat{G})|_{\text{symm}}$.  Pontryagin duality thus gives a map taking any class in $H^2(G,K)|_{\text{symm.}}$ to a class in $H^2(\hat{K},\hat{G})|_{\text{symm.}}$.  This is in fact an isomorphism, since we can always dualize again to recover the original group $\Gamma$, and thus the original extension class.\\

Dropping the assumption that the groups involved are abelian, let us comment on the meaning of (\ref{dual_ext}) in general.  (\ref{dual_ext}) is a short exact sequence of fusion categories, so in order to treat it with any mathematical rigor one requires an appropriate notion of homomorphisms between fusion categories.  Fortunately, \cite{BruguieresNatale} develops such notions (see also \cite{ENO_extensions}), and indeed proves that the short exact sequence (\ref{gamma_ext}) of groups implies the existence of the short exact sequence (\ref{dual_ext}) of fusion categories.  Appendix~\ref{appendix1} provides some examples of the sort of maps that appear in fusion categorical extensions, as opposed to group extensions.

Furthermore, the physical argument above should extend beyond the case of 0-form symmetries, meaning that any gaugeable symmetry $\Gamma=K_{[q]}.G_{[p]}$ should have a dual symmetry given by
\be
\tilde{\Gamma} = \Rep(G)_{[d-(p+2)]}.\Rep(K)_{[d-(q+2)]}
\ee
where, as explained above in section~\ref{sec:conventions}, the notion of a non-trivial extension is given by a sort of mixed anomaly generalizing the group-theoretical extension class.

\subsection{Review of the Trivializing Extension}
\label{sec:trivext}

As stated above, we are interested in the procedure of anomaly resolution via group extension as laid out in \cite{WWW}.  Specifically, we will outline the results of section 2.7 of \cite{Tachikawa}, which presents a clean mathematical version of the construction.  Assume that we have a theory in $d$ spacetime dimensions with a 0-form symmetry $G$, having non-trivial anomaly described by $\omega\in H^{d+1}(G,U(1))$.  We would like to find a finite abelian group $K$ and a short exact sequence
\be
1 \lrr K \xrightarrow{\hspace{0.2cm}\io\hspace{0.2cm}} \Gamma \xrightarrow{\hspace{0.2cm}\pi\hspace{0.2cm}} G \lrr 1
\ee
such that $\pi^*\omega$ is trivial in $H^{d+1}(\Gamma,U(1))$.

Any particular extension comes with an extension class, that is a cochain $c\in Z^2(G,K)$ -- cohomologous extension classes define isomorphic groups, so $c$ is relevant up to its class in $H^2(G,K)$.  In order to guarantee that the pullback of $\omega$ is trivial, we additionally require an element $B\in H^{d-1}(G,\hat{K})$.  $c$ and $B$ can naturally combine via the cup product to give an element of $H^{d+1}(G,U(1))$, and perhaps unsurprisingly the condition that $\pi^*\omega$ trivializes turns out to be
\be
\label{cBo}
c\cup B=\omega.
\ee

In order to construct an anomaly resolving extension for any $(G,\omega)$, then, it is sufficient to identify a finite abelian group $K$, an extension class $c\in H^2(G,K)$ and a $B\in H^{d-1}(G,\hat{K})$ satisfying (\ref{cBo}).  \cite{Tachikawa} constructs such a triple $(K,c,B)$ for arbitrary $(G,\omega)$.\footnote{It is interesting to ask, for a given pair $(G,\omega)$, what is the minimal $|K|$ required to resolve the anomaly.  The procedure of \cite{Tachikawa} results in $|K|\sim (|G|-1)^2$, but in specific examples one finds that much smaller extensions suffice.  See \cite{AEO} for a partial refinement of this construction and further discussion.}\\

Our physical interpretation of this result will be that the introduction of the trivially-acting symmetry $K$ causes the total symmetry of the theory to be
\be
(K_{[0]}.G_{[0]}).K_{[1]}=\Gamma_{[0]}.K_{[1]}
\ee
where the 1-form symmetry $K_{[1]}$ is given by the weight zero $K$ twist operators -- that is, non-trivial topological operators of codimension 2 that sit at the junction of two $K$ symmetry operators (which are of codimension 1).  The notation $\Gamma_{[0]}.K_{[1]}$ reflects that the codimension 2 $K_{[1]}$ operators are bound to the codimension 1 $\Gamma_{[0]}$ operators, and thus cannot stand on their own.  We interpret this restricted mobility of the operators as a sort of generalized group extension.

The cocycle $c$ gives the extension class for the extension of $G_{[0]}$ by $K_{[0]}$.  The element $B\in H^{d-1}(G,\hat{K})$ will be regarded as giving a mixed anomaly between $G_{[0]}$ and $K_{[1]}$.  This means that while $G_{[0]}$ is bound up in the extension with $K_{[0]}$, there should be a mixed anomaly between $\Gamma_{[0]}$ and $K_{[1]}$ given by the pullback $\pi^*B\in H^{d-1}(\Gamma,\hat{K})$.

\section{Anomaly Resolution by Group Extension}
\label{sec:2d}

We would like to examine the above formalism in more detail, beginning in two spacetime dimensions.  For concreteness, we will begin with the simplest possible example: an anomalous $\Z_2$ symmetry in 2d CFT.  This will be followed by an anomalous $\Z_2\times\Z_2$ symmetry, which allows for non-trivial non-anomalous subgroups to enter the story, after which we will turn to a general discussion of the 2d case.

Before beginning, it will behoove the reader to recall the behavior of 1-form symmetries in 2d \cite{DecompIntro}.  That is, a 1-form symmetry in two dimensions is described by a non-trivial spectrum of local (pointlike) topological operators -- usually the identity operator (corresponding to the vacuum state) would be the sole local topological operator.  This multiplicity signals the presence of a ground state degeneracy in such a theory.  Indeed, there will always be a choice of basis for these local operators for which their fusion rules become orthonormal projection.  The resulting projection operators each correspond (by state-operator correspondence) to a vacuum.  That is, the resulting theory is locally equivalent to a disjoint sum (with only extended operators being able to potentially act between separate components) \cite{Wang_2018}.

\subsection{2d Example: Anomalous $\Z_2$ Extended to $\Z_4$}
\label{sec:z22d}

Assume that we have a 2d CFT, call it $T$, with a $\Z_2$ global symmetry.  We wish to gauge (more often called orbifolding in this context) that $\Z_2$.  We will begin with an analysis at the level of partition functions, demonstrating how an 't Hooft anomaly manifests as an obstruction to modular invariance and how a group extension can cure this, as examined previously in \cite[section 7.1]{Wang_2018} and \cite[section 5.1.1]{AEO}.  Afterwards we will re-examine the situation more abstractly in terms of symmetries.

\subsubsection{Partition Function Analysis}
\label{sec:z2pf}

Let $Z^T$ be the torus partition function of $T$.  The partition function of the orbifold theory breaks into sectors $Z^T_{g_1,g_2}$ in which we wrap the cycles of the torus with line operators corresponding to $g_1$ and $g_2$.  In path integral language, this amounts to considering group-twisted boundary conditions along the cycles.  Either way, the partition function of the orbifold theory will be
\be
\label{z2orb}
Z^{T/\Z_2}=\frac{1}{2}[Z^T_{0,0}+Z^T_{0,1}+Z^T_{1,0}+Z^T_{1,1}]
\ee
in which we have written $\Z_2$ additively as $\{0,1\}$ and normalized by the order of the gauge group.

Now let us see how the presence of an 't Hooft anomaly would disrupt such a procedure.  We get away with having only four sectors in (\ref{z2orb}) because any configuration of lines wrapping the cycles of the torus can be reduced to one of those four.  For instance, consider the configuration implied by $Z^T_{2,0}$, in which one of the cycles has been wrapped with two non-trivial $\Z_2$ lines.  We would simply fuse these lines to get the identity and conclude that $Z^T_{2,0}$ was redundant with $Z^T_{0,0}$.

But we are being too fast if we assume that we can always do this -- reducing an arbitrary configuration to one of the four appearing in (\ref{z2orb}) may require resolving four-way junctions into three-way junctions (and similarly for higher junctions).  As discussed earlier, the sort of anomalies under consideration manifest as a phase ambiguity in such a resolution.  In fact, we can tie such a discussion to the modular properties of the $Z_{g_1,g_2}$.  For $g$ an element of order $n$, we have the following modular transformation \cite{AEO}:
\be
\label{anomt}
Z_{g,g'}(\tau+n)=Z_{g,g'}(\tau) \prod_{i=1}^n\omega(g^{-1},g^i,g)
\ee
where $\omega\in H^3(G,U(1))$ (the product appearing above is invariant under coboundary shifts) classifies the 't Hooft anomaly in $G$.

In the example at hand of a $\Z_2$ symmetry, we have $H^3(\Z_2,U(1))=\Z_2$, so there is a single type of non-trivial anomaly that could appear.  In the presence of such an anomaly, we would have e.g.
\be
Z^T_{3,1}(\tau)=Z^T_{1,1}(\tau+2)=-Z^T_{1,1}(\tau).
\ee
This is a problem.  We would hope that our orbifold theory is a well-defined CFT, and so it ought to exhibit modular invariance.  When the $\Z_2$ symmetry in question is anomalous, however, applying (\ref{anomt}) to the orbifold partition function (\ref{z2orb}) gives transformations like
\begin{multline}
Z^{T/\Z_2}(\tau+2)=\frac{1}{2}[Z^T_{0,0}(\tau+2)+Z^T_{0,1}(\tau+2)+Z^T_{1,0}(\tau+2)+Z^T_{1,1}(\tau+2)]\\
=\frac{1}{2}[Z^T_{0,0}+Z^T_{0,1}-Z^T_{1,0}-Z^T_{1,1}].
\end{multline}
Thus the 't Hooft anomaly disrupts modular invariance of the putative orbifold theory (it ought to be, in the absence of an anomaly, that modular transformations simply permute terms in the sum of (\ref{z2orb}), leaving the end result invariant).\\

Now we wish to remedy this issue by the group extension procedure laid out in section~\ref{sec:trivext}.  One can readily show that extending our $\Z_2$ symmetry to $\Z_4$ will be sufficient for $\omega$ to pull back to the trivial class.  The extension class $c$ is the non-trivial element of $H^2(\Z_2,\Z_2)=\Z_2$ and the mixed anomaly $B$ is the non-trivial class in $H^1(\Z_2,\Z_2)=\Z_2$.  The claim, then, is that the resulting $\Z_4$ symmetry is non-anomalous and can safely be gauged.  The $\Z_4$ partition function would take the form
\be
\label{z4orb}
Z^{T/\Z_4}=\frac{1}{4}\sum_{i,j=0}^3Z^T_{i,j}.
\ee
Since the $\Z_2$ subgroup of $\Z_4$ has trivial action on the theory, we should be able to equate the sectors appearing in this $\Z_4$ orbifold to those of the original $\Z_2$ orbifold in (\ref{z2orb}).  This procedure is almost as simple as taking the indices appearing in (\ref{z4orb}) mod 2 -- however, there is one ingredient we have not yet mentioned.  

The mixed anomaly $B$, which was purported as being important to the anomaly resolution procedure, has yet to appear.  $B$ will in fact contribute relative phases between the different $\Z_4$ sectors as we reduce them to $\Z_2$ sectors.  It's a good thing, too, as if we simply took the indices on the $\Z_4$ sectors mod 2, we would find that (\ref{z4orb}) becomes two copies of (\ref{z2orb}), and the anomaly would persist.  Explicitly, taking the mixed anomaly phases into account gives us
\begin{align}
\label{z4red}
Z^{T/\Z_4}_{0,0}&=Z^{T/\Z_2}_{0,0},\\\notag
Z^{T/\Z_4}_{0,1}&=Z^{T/\Z_2}_{0,1},\\\notag
Z^{T/\Z_4}_{0,2}&=Z^{T/\Z_2}_{0,0},\\\notag
Z^{T/\Z_4}_{0,3}&=Z^{T/\Z_2}_{0,1},\\\notag
Z^{T/\Z_4}_{1,0}&=Z^{T/\Z_2}_{1,0},\\\notag
Z^{T/\Z_4}_{1,1}&=Z^{T/\Z_2}_{1,1},\\\notag
Z^{T/\Z_4}_{1,2}&=-Z^{T/\Z_2}_{1,0},\\\notag
Z^{T/\Z_4}_{1,3}&=-Z^{T/\Z_2}_{1,1},\\\notag
Z^{T/\Z_4}_{2,0}&=Z^{T/\Z_2}_{0,0},\\\notag
Z^{T/\Z_4}_{2,1}&=-Z^{T/\Z_2}_{0,1},\\\notag
Z^{T/\Z_4}_{2,2}&=Z^{T/\Z_2}_{0,0},\\\notag
Z^{T/\Z_4}_{2,3}&=-Z^{T/\Z_2}_{0,1},\\\notag
Z^{T/\Z_4}_{3,0}&=Z^{T/\Z_2}_{1,0},\\\notag
Z^{T/\Z_4}_{3,1}&=-Z^{T/\Z_2}_{1,1},\\\notag
Z^{T/\Z_4}_{3,2}&=-Z^{T/\Z_2}_{1,0},\\\notag
Z^{T/\Z_4}_{3,3}&=Z^{T/\Z_2}_{1,1}.\\\notag
\end{align}
Combining (\ref{z4red}) and (\ref{z4orb}), we find that
\be
\label{z4gaugeresult}
Z^{T/\Z_4}=Z^T_{0,0}=Z^T.
\ee
That is, the gauged theory is equivalent to the one we began with.  All of the anomalous contributions have dropped out as well -- indeed, modular invariance of the orbifold theory is tautological.

\subsubsection{Analysis in Terms of Symmetries}
\label{sec:z2sym}

Let us repeat the above example, giving a careful accounting of the topological operators which appear at each step.  Our theory initially has the zero-form symmetry $G_{[0]}=\Z_2$.  As discussed in section~\ref{sec:trivext}, the addition of the trivially-acting symmetry $K$ should be interpreted as giving us a total symmetry
\be
\label{z2resolved}
(K_{[0]}.G_{[0]}).K_{[1]}=((\Z_2)_{[0]}.(\Z_2)_{[0]}).(\Z_2)_{[1]}=(\Z_4)_{[0]}.(\Z_2)_{[1]}.
\ee
Let us gauge $K_{[0]}$.  The analysis of section~\ref{sec:extswap} now comes into play, as we are gauging a symmetry that is participating in extensions.  Both the $G_{[0]}$ lines and the $K_{[1]}$ points were bound to $K_{[0]}$ and did not constitute standalone subsymmetries -- gauging $K_{[0]}$ will free them, at the cost of introducing mixed anomalies.  The symmetry of the gauged theory should then be
\be
\label{z2gauge_ex_sym}
(\hat{\Z}_2)_{[0]}\times(\Z_2)_{[0]}\times(\Z_2)_{[1]},
\ee
and we see that the resulting symmetry is given quite simply by direct products.  While this is an authentic presentation of the global symmetry of the gauged theory, keep in mind that this notation does not capture the mixed anomalies present.  From the beginning, there was a mixed anomaly, given by $B$, between $G_{[0]} = (\Z_2)_{[0]}$ and $(\Z_2)_{[1]}$.  Furthermore, by gauging $K_{[0]}$ we have turned the extension class between it and $G_{[0]}$ into a mixed anomaly, as well as the extension class between it and $K_{[1]}$.  The upshot of this is that any two terms in (\ref{z2gauge_ex_sym}) have a mixed anomaly between them.  Said another way, we could gauge the resulting theory by $(\hat{\Z}_2)_{[0]}$, $(\Z_2)_{[0]}$ or $(\Z_2)_{[1]}$, but not any mix thereof.

The previous statement may be somewhat surprising -- the theory in question appears to have $G_{[0]}=(\Z_2)_{[0]}$ as a symmetry, and we have just claimed that it is non-anomalous (by virtue of the only anomalies being mixed ones).  But the whole premise was that the symmetry $G_{[0]}$ was anomalous, so what gives?  The answer will be somewhat subtle -- the $(\Z_2)_{[0]}$ appearing as a symmetry of this gauged theory isn't \textit{quite} the one we started with, though they're certainly related.  Analyzing the role of the one-form symmetry will help to clarify this point.\\

Let us begin by giving the topological operators that implement $(\Z_2)_{[1]}$ a name -- call them $\sigma_1$ and $\sigma_k$.  The meaning behind calling this a $\Z_2$ 1-form symmetry is that these operators must fuse according to the $\Z_2$ group law:
\be
\label{sigma_fuse}
\sigma_1\otimes\sigma_1=\sigma_1,\hspace{0.5cm}\sigma_1\otimes\sigma_k=\sigma_k,\hspace{0.5cm}\sigma_k\otimes\sigma_1=\sigma_k,\hspace{0.5cm}\sigma_k\otimes\sigma_k=\sigma_1.
\ee
These are local operators which we can regard as living in a Hilbert space; that is, we can take arbitrary complex combinations of the $\sigma_i$.  Let us define
\be
\pi_\pm=\frac{1}{2}(\sigma_1\pm\sigma_g).
\ee
From (\ref{sigma_fuse}) one immediately sees that the $\pi_\pm$ fuse as
\be
\pi_+\otimes\pi_+=\pi_+,\hspace{0.5cm}\pi_-\otimes\pi_-=\pi_-,\hspace{0.5cm}\pi_+\otimes\pi_-=\pi_-\otimes\pi_+=0,
\ee
such that they form orthonormal projectors.  Physically, $\pi_+$ and $\pi_-$ form vacua for two disjoint copies of the theory in question, and we should regard the theory gauged by $K_{[0]}$ as being such a disjoint union.\\

Importantly, both 0-form symmetries will have non-trivial action on $\pi_\pm$.  The quantum symmetry $(\hat{\Z}_2)_{[0]}$ is in fact defined by its action on the $\sigma_i$, which were the $K_{[0]}$ twist fields -- it acts by characters.  In this example, if $\mcL_{\hat{k}}$ denotes the non-trivial line operator of $(\hat{\Z}_2)_{[0]}$, we should have
\be
\mcL_{\hat{k}}\cdot\sigma_1=\sigma_1\hspace{0.5cm}\text{and}\hspace{0.5cm}\mcL_{\hat{k}}\cdot\sigma_k=-\sigma_k.
\ee
Translating this into an action on $\pi_\pm$, we have
\be
\mcL_{\hat{k}}\cdot\pi_+=\pi_-\hspace{0.5cm}\text{and}\hspace{0.5cm}\mcL_{\hat{k}}\cdot\pi_-=\pi_+.
\ee
That is, the action of the dual of a trivially-acting symmetry is to exchange the disjoint copies that resulted from gauging it.  Hence the adage that in 2d, trivially-acting symmetries are dual to exchange symmetries.\\

Now we would like to consider the action of $G_{[0]}=(\Z_2)_{[0]}$ on the $\sigma_i$ and thus $\pi_\pm$.  This is straightforward -- the mixed anomaly between $G_{[0]}$ and $K_{[1]}$ can be interpreted as a phase that arises when a $G$ line wraps a $K$ point.  Our non-trivial mixed anomaly $B$ corresponds to the action, if $\mcL_g$ is the line associated with the non-trivial element of $G_{[0]}$,
\be
\mcL_g\cdot\sigma_1=\sigma_1\hspace{0.5cm}\text{and}\hspace{0.5cm}\mcL_g\cdot\sigma_k=-\sigma_k.
\ee
This is, of course, the exact same action on the $\sigma$ as $\mcL_{\hat{k}}$, so we find the same action on the $\pi$ as well:
\be
\mcL_g\cdot\pi_+=\pi_-\hspace{0.5cm}\text{and}\hspace{0.5cm}\mcL_g\cdot\pi_-=\pi_+.
\ee

In the theory gauged by $K_{[0]}$, then, both $\hat{K}_{[0]}$ and $G_{[0]}$ have the effect of swapping the vacua $\pi_\pm$.  The symmetry that acts as $G$ originally did on both components is now the diagonal action of $\hat{K}_{[0]}\times G_{[0]}$ -- and appropriately, that action is anomalous, captured by the mixed anomaly between $\hat{K}_{[0]}$ and $G_{[0]}$.  Effectively, by adding a second copy of the theory, we have been able to take our anomalous $\Z_2$ symmetry and construct another symmetry, also given by the group $\Z_2$, which is non-anomalous because it mixes the original $\Z_2$ action with a swap action on the copies.\\

Finally, as the $(\Z_2)_{[0]}$ appearing in (\ref{z2gauge_ex_sym}) is non-anomalous, we could consider gauging it.  As before, gauging will swap the mixed anomalies in which $(\Z_2)_{[0]}$ participates for extension classes, and we would expect to find a theory whose symmetry is
\be
((\hat{\Z}_2)_{[0]}.(\hat{\Z}_2)_{[0]}).(\Z_2)_{[1]}=(\hat{\Z}_4)_{[0]}.(\Z_2)_{[1]}.
\ee
Now, in comparison to when we started, the one-form symmetry $K_{[1]}=(\Z_2)_{[1]}$ is bound by extension class to $\hat{G}_{[0]}=(\hat{\Z}_2)_{[0]}$, which means that the resulting $\hat{G}$ symmetry is trivially-acting.  The total symmetry is an extension of the effective $\hat{K}$ symmetry by the trivially-acting $\hat{G}$; $G$ and $K$ have truly swapped roles.  Because the 1-form symmetry is bound to the $\hat{G}$ lines, the resulting theory once again contains only a single vacuum.  This exactly matches the partition function analysis of section~\ref{sec:z2pf}, where in (\ref{z4gaugeresult}) we found that the $(\Z_4)_{[0]}$ orbifold partition function, accounting for the mixed anomaly, produces a single copy of the parent theory partition function.

\subsection{2d Example: Anomalous $\Z_2\times\Z_2$ Extended to $Q_8$}
\label{sec:z2z22d}

Despite the simplicity of the previous example, it may have been difficult to disentangle the roles of $G$ and $K$, as they were both $\Z_2$ symmetries.  Additionally, $G$ had no non-anomalous subgroups, a feature which we will see can qualitatively alter results.  Accordingly, we now move on to the next-simplest example, in which $G=\Z_2\times\Z_2$ with a single non-anomalous $\Z_2$ subgroup.  We can once again take $K$ to be $\Z_2$, with $\Gamma$ being $Q_8$, the group of unit quaternions.\footnote{This example appeared in \cite[appendix A]{AEO}.  See also \cite[section 7.2]{Wang_2018} for a similar scenario involving an extension to the dihedral group $D_4$.}  The symmetry of our initial theory is then
\be
((\Z_2)_{[0]}.(\Z_2\times\Z_2)_{[0]}).(\Z_2)_{[1]}=(Q_8)_{[0]}.(\Z_2)_{[1]}.
\ee

As before, we begin by gauging $K_{[0]}=(\Z_2)_{[0]}$, which brings us to a theory with symmetry
\be
(\hat{\Z}_2)_{[0]}\times(\Z_2\times\Z_2)_{[0]}\times(\Z_2)_{[1]}
\ee
with both extension classes turned into mixed anomalies.  $(\Z_2\times\Z_2)_{[0]}$ will now include a swap as part of its action, and again there will be a $\Z_2\times\Z_2$ subsymmetry of $(\hat{\Z}_2)_{[0]}\times(\Z_2\times\Z_2)_{[0]}$ that replicates the original anomaly.  Gauging $(\Z_2\times\Z_2)_{[0]}$ results in a theory with symmetry
\be
((\hat{\Z}_2\times\hat{\Z}_2)_{[0]}.(\hat{\Z}_2)_{[0]}).(\Z_2)_{[1]}=\Rep(Q_8)_{[0]}.(\Z_2)_{[1]},
\ee
which is the same theory we would have obtained via gauging the entire $Q_8$ symmetry of the original theory.

Note that the $\hat{G}$ symmetry, $(\hat{\Z}_2\times\hat{\Z}_2)_{[0]}$, has the $(\Z_2)_{[1]}$ bound to it -- this tells us that part of this symmetry is trivially-acting.  Indeed, in contrast to the previous example, in which the $\hat{G}$ symmetry was entirely trivially-acting, this example has $\hat{G}$ built out of a \textit{mix} of a trivially-acting symmetry and (the dual to) a non-anomalous subgroup of $G$.  That is, the result of gauging the resolution of the anomalous $\Z_2\times\Z_2$ symmetry is a dual $\hat{\Z}_2\times\hat{\Z}_2$ built as a direct product of the $\hat{\Z}_2$ dual to the non-anomalous subgroup of the original and a trivially-acting $\Z_2$.  Again, the anomaly resolution procedure has `surgically' removed the anomalous part of the group, replacing it with a swapping action (the dual to which is trivially-acting).

\subsection{General Case in 2d}

With two examples under our belt, let us state the general case for resolving an anomalous 2d symmetry.  We begin with an anomalous 0-form symmetry $G_{[0]}$, with anomaly specified by $\omega$ representing a non-trivial class in $H^3(G,U(1))$.  We choose a finite abelian group $K$ along with a $c\in H^2(G,K)$ and $B\in H^1(G,\hat{K})$ satisfying $\omega=c\cup B$.  We extend $G_{[0]}$ by a trivially-acting $K$ symmetry; that is, we interpret our theory as having symmetry
\be
(K_{[0]}.G_{[0]}).K_{[1]}=\Gamma_{[0]}.K_{[1]}
\ee
where $K_{[1]}$ are the weight zero twist operators that live at junctions between lines which differ by an element of $K$.  $c$ provides the extension class for $\Gamma$, while $B$ gives the mixed anomaly between the $G_{[0]}$ lines and the $K_{[1]}$ points.\\

Now we gauge $K_{[0]}$.  The resulting theory has symmetry
\be
\hat{K}_{[0]}\times G_{[0]}\times K_{[1]}
\ee
with three mixed anomalies.  $B$, as stated above, provides a mixed anomaly between $G_{[0]}$ and $K_{[1]}$.  The $\hat{K}_{[0]}$ lines act by characters on the $K_{[1]}$ points, which provides a mixed 0/1-form anomaly between the two.  Finally, there is a mixed $\hat{K}_{[0]}/G_{[0]}$ anomaly.  Following the discussion in \cite[section 2.2]{Tachikawa}, one can see this mixed anomaly as in Figure~\ref{fig:kgma}.\\

\begin{figure}
\begin{subfigure}{0.5\textwidth}
\centering
\begin{tikzpicture}
\draw[very thick,->] (-2,-1) -- (-1,-0.5);
\draw[very thick] (-1,-0.5) -- (0,0);
\node at (-1,-1) {$(1,g_1)$};
\draw[very thick,->] (2,-1) -- (1,-0.5);
\draw[very thick] (1,-0.5) -- (0,0);
\node at (1,-1) {$(1,g_2)$};
\filldraw[black] (0,0) circle (2pt);
\node at (1,0) {$\sigma_{c(g_1,g_2)}$};
\draw[very thick,->] (0,0) -- (0,1);
\draw[very thick] (0,1) -- (0,2);
\node at (1.5,1.5) {$(c(g_1,g_2),g_1g_2)$};
\end{tikzpicture}
\caption{}
\label{kgma1}
\end{subfigure}
\begin{subfigure}{0.5\textwidth}
\centering
\begin{tikzpicture}
\draw[very thick,->] (0,0) [partial ellipse=0:-180:1cm and 1cm];
\draw[very thick,->] (0,0) [partial ellipse=-180:-360:1cm and 1cm];
\node at (0,1.5) {$\hat{k}$};
\filldraw[black] (0,0) circle (2pt);
\node at (0,-0.3) {$\sigma_k$};
\node at (1.5,0) {$\Rightarrow$};
\filldraw[black] (2,0) circle (2pt);
\node at (2,-0.3) {$\sigma_k$};
\node at (3,0) {$\times\text{ }\chi_{\hat{k}}(k)$};
\end{tikzpicture}
\caption{}
\label{kgma2}
\end{subfigure}
\begin{subfigure}{0.5\textwidth}
\centering
\begin{tikzpicture}
\draw[very thick,->] (-2,-1) -- (-1,-0.5);
\draw[very thick] (-1,-0.5) -- (0,0);
\node at (-1,-1) {$g_1$};
\draw[very thick,->] (2,-1) -- (1,-0.5);
\draw[very thick] (1,-0.5) -- (0,0);
\node at (1,-1) {$g_2$};
\filldraw[black] (0,0) circle (2pt);
\draw[very thick,->] (0,0) -- (0,1.5);
\draw[very thick] (0,1.5) -- (0,2);
\node at (0.75,1.5) {$g_1g_2$};
\draw[very thick,->] (-2,0.75) -- (-1,0.75);
\draw[very thick,->] (-1,0.75) -- (1,0.75);
\draw[very thick] (1,0.75) -- (2,0.75);
\node at (-1,1.25) {$\hat{k}$};
\end{tikzpicture}
\caption{}
\label{kgma3}
\end{subfigure}
\begin{subfigure}{0.5\textwidth}
\centering
\begin{tikzpicture}
\draw[very thick,->] (-2,-1) -- (-0.5,-0.25);
\draw[very thick] (-0.5,-0.25) -- (0,0);
\node at (-1.1,-0.25) {$g_1$};
\draw[very thick,->] (2,-1) -- (0.5,-0.25);
\draw[very thick] (0.5,-0.25) -- (0,0);
\node at (1.1,-0.25) {$g_2$};
\filldraw[black] (0,0) circle (2pt);
\draw[very thick,->] (0,0) -- (0,1);
\draw[very thick] (0,1) -- (0,2);
\node at (0.75,1.5) {$g_1g_2$};
\draw[very thick,->] (-2,-0.75) -- (0,-0.75);
\draw[very thick] (0,-0.75) -- (2,-0.75);
\node at (0,-1.25) {$\hat{k}$};
\node at (2.5,0.5) {$\times$};
\node at (4,0.5) {$\chi_{\hat{k}}(c(g_1,g_2))$};
\end{tikzpicture}
\caption{}
\label{kgma4}
\end{subfigure}
\caption{A visualization of the mixed anomaly between $\hat{K}_{[0]}$ and $G_{[0]}$.}
\label{fig:kgma}
\end{figure}

In Figure~\ref{kgma1}, we represent a group extension as a choice of junction operator between $G$ lines.  That is, given two lines labeled by elements $g_1$ and $g_2$ of $G$, an extension of $G$ to $\Gamma=K.G$ with extension class $c\in H^2(G,K)$ can be visualized as the insertion of a local operator labeled by $\sigma_{c(g_1,g_2)}$.  The choice of a cocycle representative for $c$ is equivalent to choosing an identification of the elements of $\Gamma$ with those of $K\times G$, allowing us to label a $\Gamma$ line by a pair $(k,g)$.  Then, as in the figure, when a $(1,g_1)$ line meets and fuses with a $(1,g_2)$ line, the resulting line is labeled by $(c(g_1,g_2),g_1g_2)$.

We will be interested in the action of a $\hat{K}$ line on the (local) junction operator.  $\hat{K}$ is defined through its action on $K$ twist fields, as shown in Figure~\ref{kgma2}.  That is, a $\hat{k}$ line wrapping a local operator $\sigma_k$ can be shrunk to reobtain $\sigma_k$ multiplied by the phase $\chi_{\hat{k}}(k)$.

Now we would like to use this information to ask about the interaction of $\hat{K}$ lines with $G$ lines in the gauged theory.  Figure~\ref{kgma3} shows a $\hat{k}$ line about to cross a three-way junction of $G$ lines (in the gauged theory $G_{[0]}$ is a standalone symmetry, so the lines are once again labeled purely by elements of $G$).  Pulling the $\hat{k}$ line down through the intersection has the potential to generate a phase.  Indeed, if we consider lifting this configuration to the ungauged theory, we would find a phase difference between Figure~\ref{kgma3} and Figure~\ref{kgma4} given by the action of the $\hat{k}$ line on the $\sigma_{c(g_1,g_2)}$ junction operator.  That is, the phase we obtain in dragging the $\hat{k}$ line across the junction should be $\chi_{\hat{k}}(c(g_1,g_2))$.  This provides a mixed anomaly in the gauged theory between $\hat{K}_{[0]}$ and $G_{[0]}$, our third mixed anomaly (of three possible combinations).\\

As in the examples of sections~\ref{sec:z22d} and \ref{sec:z2z22d}, the mixed anomalies of $\hat{K}_{[0]}$ and $G_{[0]}$ with $K_{[1]}$ will guarantee that both of the 0-form symmetries have the swapping of vacua as part of their action.  The anomaly that originally prevented us from gauging $G_{[0]}$ is compensated by its new swapping behavior, and this anomaly has essentially been `shifted' into the mixed anomaly between $\hat{K}_{[0]}$ and $G_{[0]}$.  We can see this by defining a new $G$ symmetry, $\tilde{G}_{[0]} \in \hat{K}_{[0]}\times G_{[0]}$, whose lines are labeled by $(\hat{k},g)=(B(g,k),g)$ (recalling that for each $g\in G$, $B(g,k)$ gives a $K$ character).  The calculations of Figure~\ref{fig:kgma} show that $\tilde{G}_{[0]}$ has the anomaly $B(g_3,c(g_1,g_2))$.  This is $c\cup B$, which by design is equal to the original $G$ anomaly $\omega$.  Thus, the original anomaly manifests in a subsymmetry of $\hat{K}_{[0]}\times G_{[0]}$.\\

Finally, we gauge $G_{[0]}$.  This brings us to a theory with symmetry
\be
(\Rep(G)_{[0]}.\hat{K}_{[0]}).K_{[1]}=\Rep(\Gamma)_{[0]}.K_{[1]},
\ee
which is of course the same theory we would have found had we gauged the entire anomaly-resolved $\Gamma_{[0]}$ symmetry of the initial theory.  The two gaugings have effected an extension swap -- the resulting symmetry is an extension of $\Rep(K)=\hat{K}$ by $\Rep(G)$.

The 1-form symmetry $K_{[1]}$ is bound to the $\Rep(G)_{[0]}$ lines, signaling that at least part of that symmetry is trivially-acting.  Following the $\Z_2\times\Z_2$ example of section~\ref{sec:z2z22d}, we should expect that $\Rep(G)_{[0]}$ is built out of a mix of symmetries dual to the non-anomalous subgroups of $G$ and trivially-acting symmetries.

In the case that $G$ is abelian, we can readily see how the mixed anomaly $B$ between $G_{[0]}$ and $K_{[1]}$ determines an extension class, which we would expect to be valued in $H^1(K,\hat{G})$ \cite{lfs}.  In this situation $B$ is a map $G\times K\to U(1)$, so we simply regard it as producing a $G$ character for each element of $K$.  In a more general situation we would expect $B$ to determine a map in $\Hom{(K,\Rep(G))}$, though this scenario takes us beyond group cohomology.

\subsection{3d and Beyond}
\label{sec:3d}

Let us move on to the case of a 3d QFT.  Once again we write our anomalous 0-form symmetry as $G_{[0]}$, with anomaly given by $\omega\in H^4(G,U(1))$.  We introduce a trivially-acting finite abelian 0-form symmetry $K$ such that the total symmetry is
\be
(K_{[0]}.G_{[0]}).K_{[1]}.
\ee
This requires specifying an extension class $c\in H^2(G,K)$ for the 0-form symmetries and a mixed anomaly $B\in H^2(G,\hat{K})$ between $G_{[0]}$ and $K_{[1]}$, satisfying $c\cup B = \omega$.

Gauging $K_{[0]}$ brings us to a theory with symmetry
\be
\hat{K}_{[1]}\times G_{[0]} \times K_{[1]}
\ee
and mixed anomalies between any two terms.  Specifically, $B$ still provides the mixed anomaly between $G_{[0]}$ and $K_{[1]}$.  $c$ clearly specifies an element of $H^2(G,\hat{\hat{K}})$, and thus describes a mixed anomaly between $G_{[0]}$ and $\hat{K}_{[1]}$.  $\hat{K}_{[1]}$ acts on $K_{[1]}$ via characters, as is usual for a quantum symmetry, and this action provides a mixed anomaly between the two.

As in the 2d case, the $G_{[0]}$ symmetry appearing in the theory gauged by $K_{[0]}$ will be non-anomalous.  This is possible because it has, via mixed anomaly, an action on the $K_{[1]}$ (and $\hat{K}_{[1]}$) lines.  Again there will be a subsymmetry of $\hat{K}_{[1]}\times G_{[0]}$ which replicates the original, anomalous $G$ symmetry.  This symmetry is essentially $G_{[0]}$, except at the codimension 2 intersection between two planes $g_1$ and $g_2$ we insert the $\hat{K}_{[1]}$ line given by $B(g_1,g_2,k)$.

Finally, we can gauge $G_{[0]}$ to arrive at a theory with symmetry
\be
\label{3dresult}
(\Rep(G)_{[1]}.\hat{K}_{[1]}).K_{[1]}.
\ee
Three dimensions is notable because the symmetry of the resulting theory is a pure 1-form symmetry.  For instance, if $G$ and $\Gamma=K.G$ are abelian, (\ref{3dresult}) becomes
\be
\label{3dresultab}
(\hat{G}_{[1]}.\hat{K}_{[1]}).K_{[1]}=\hat{\Gamma}_{[1]}.K_{[1]},
\ee
which is a group-like one-form symmetry.  Specifically, Pontryagin duality allows us to map the original extension class $c\in H^2(G,K)$ to $\tilde{c}\in H^2(\hat{K},\hat{G})$, which provides an extension class for $\Gamma$.  Similarly, the mixed anomaly between $G_{[0]}$ and $K_{[1]}$ is given by $B\in H^2(G,\hat{K})$.  Again Pontryagin duality allows us to map this to $\tilde{B}\in H^2(K,\hat{G})$ -- we then use inclusion of $\hat{G}$ in $\hat{\Gamma}$ to get an element of $H^2(K,\hat{\Gamma})$ which provides the second extension class.  Clearly it is the fact that, in 3d, the mixed 0/1-form anomaly is specified by second cohomology that allows for the resulting symmetry to be purely group-like.\\

At this point, increasing the dimension further will not appreciably change the story.  2d was special due to the presence of a $(d-1)$-form symmetry and therefore multiple vacua, but higher dimensions will resemble the 3d story presented above.  We will extend $G_{[0]}$ by the trivially-acting $K_{[0]}$, imposing a mixed anomaly between $G_{[0]}$ and the weight zero $K$ twist fields that generate $K_{[1]}$.  This initial theory has symmetry
\be
(K_{[0]}.G_{[0]}).K_{[1]}=\Gamma_{[0]}.K_{[1]}.
\ee
Gauging $K_{[0]}$ brings us to a theory with a non-anomalous $G_{[0]}$ symmetry -- the original, anomalous $G$ action will now be given by a subsymmetry of $\hat{K}_{[d-2]}\times G_{[0]}$.  The total symmetry is the direct product
\be
\hat{K}_{[d-2]}\times G_{[0]}\times K_{[1]}
\ee
with maximally mixed anomaly.  Gauging $G_{[0]}$ produces a theory with symmetry
\be
(\Rep(G)_{[d-2]}.\hat{K}_{[d-2]}).K_{[1]}=\Rep(\Gamma)_{[d-2]}.K_{[1]}.
\ee

\section{Anomaly Resolution by Higher Extension}
\label{sec:1d}

The case of one spacetime dimension (quantum mechanics) is somewhat special for the anomaly resolution story as we've told it so far.  Until now, the procedure has been to extend our anomalous 0-form symmetry $G$ by a trivially-acting 0-form symmetry $K$.  We have been able to specify the anomaly's class in $H^{d+1}(G,U(1))$ by cupping the extension class with the mixed anomaly between $G$ and the 1-form symmetry that controls the triviality of $K$.  1d presents an issue for this prescription, in that there are no 1-form symmetries.  The anomaly resolution process as laid out in section~\ref{sec:trivext} still applies in 1d, but we are going to need to develop a separate interpretation.

In 1d, 0-form symmetries are controlled by local operators.  An anomaly in a symmetry $G$ is specified by a class in $H^2(G,U(1))$, and the interpretation of this is that the corresponding local operators fuse in a projective representation of $G$.  The anomaly resolution procedure we have been using so far would produce a finite abelian group $K$, an extension class $c\in H^2(G,K)$ and an element $B$ of $H^0(G,\hat{K})=\hat{K}$ -- in this dimension, $B$ is simply a $K$ character, with no $G$ dependence.  As before, we have the condition that $c \cup B=\omega\in H^2(G,U(1))$.

\subsection{1d Example: Anomalous $\Z_2\times\Z_2$ Extended to $Q_8$}

Undaunted, let us push forward with an example of resolving an anomalous symmetry in 1d.  The simplest group with a non-trivial projective representation is $\Z_2\times\Z_2$, for which we have $H^2(\Z_2\times\Z_2,U(1))=\Z_2$.  We expect to be able to resolve this anomaly with a $\Z_2$ extension.  Writing $\Z_2\times\Z_2$ as $\{1,i,j,k\}$, we can take the extension class to have non-trivial values
\be
c(i,i)=c(j,j)=c(k,k)=c(j,i)=c(k,j)=c(i,k)=-1
\ee
where the extending $\Z_2$ is being written as $\{1,-1\}$.  The full group is then $\Z_2.(\Z_2\times\Z_2)=Q_8$ with its usual presentation $\braket{i,j,k|i^2=j^2=k^2=ijk=-1}$ (we are abusing notation by using the same symbols for the generators of $Q_8$ and that of its $\Z_2\times\Z_2$ quotient).

So, beginning with an anomalous (that is, projective) $\Z_2\times\Z_2$ symmetry, we extend it to
\be
\label{1dstep1}
(\Z_2)_{[0]}.(\Z_2\times\Z_2)_{[0]}=(Q_8)_{[0]}.
\ee
Gauging the extending $\Z_2$ symmetry changes the above to
\be
\label{1dstep2}
(\hat{\Z}_2)_{[-1]}\times (\Z_2\times\Z_2)_{[0]}
\ee
where, as before, we expect the $(\Z_2\times\Z_2)_{[0]}$ symmetry of this theory to be non-anomalous.  How does the symmetry (\ref{1dstep2}) encode the original $G$ anomaly?  Recall \cite{lfs} that a $-1$-form symmetry is associated with codimension 0 topological defects, i.e.~spacetime-filling defects.  Let the non-trivial defect in $(\hat{\Z}_2)_{[-1]}$ be labeled by $\hat{k}$, which we will take to be the non-trivial $\Z_2$ character.  $\hat{k}$, being spacetime-filling, constitutes a \textit{background defect} for the theory.

\begin{figure}
\begin{subfigure}{0.5\textwidth}
\centering
\begin{tikzpicture}
\draw[very thick,->] (-2,0) -- (-1,0);
\draw[very thick] (-1,0) -- (2,0);
\node at (-1,0.5) {$\hat{k}$};
\filldraw[black] (0,0) circle (2pt);
\node at (0,-0.5) {$g_1$};
\filldraw[black] (1,0) circle (2pt);
\node at (1,-0.5) {$g_2$};
\end{tikzpicture}
\caption{}
\label{1dfuse1}
\end{subfigure}
\begin{subfigure}{0.5\textwidth}
\centering
\begin{tikzpicture}
\draw[very thick,->] (-2,0) -- (-1,0);
\draw[very thick] (-1,0) -- (2,0);
\node at (-1,0.5) {$\hat{k}$};
\filldraw[black] (0.5,0) circle (2pt);
\node at (0.5,-0.5) {$\chi_{\hat{k}}(c(g_1,g_2))g_1g_2$};
\end{tikzpicture}
\caption{}
\label{1dfuse2}
\end{subfigure}
\caption{Fusion of $G_{[0]}$ operators becomes projective in the $\hat{k}$ background.}
\label{fig:1dfuse}
\end{figure}

The resolution will be that the $(\Z_2\times\Z_2)_{[0]}$ operators fuse projectively in the $\hat{k}$ background.  This configuration is pictured in Figure~\ref{fig:1dfuse}, where Figure~\ref{1dfuse1} shows the local operators $g_1$ and $g_2$ in the presence of a background defect labeled by $\hat{k}$.  When these operators fuse, as shown in Figure~\ref{1dfuse2}, they pick up the phase $\chi_{\hat{k}}(c(g_1,g_2))$, and our extension was designed exactly such that this is equal to the original anomaly $\omega\in H^2(G,U(1))$.

\subsection{General Procedure}

The method outlined above for 1d anomaly resolution can in fact be replicated in higher dimensions, giving an alternative procedure to the one studied in section~\ref{sec:2d}, first proposed in \cite{Kobayashi_2019}.  Again we begin with a 0-form symmetry $G_{[0]}$ with anomaly $\omega\in H^{d+1}(G,U(1))$.  We extend this symmetry by a $(d-1)$-form symmetry $K_{[d-1]}$ such that we have the symmetry
\be
K_{[d-1]}.G_{[0]}.
\ee
The extension class $c$ is valued in $H^{d+1}(G,K)$ -- our choice of $K$ and $c$ is dictated by the requirement that there is a character $\hat{k}$ of $K$ such that $c\cup \hat{k}=\omega$.  The interpretation of this construction is that we begin with multiple copies of a theory (as signaled by the $(d-1)$-form symmetry).  Gauging $K_{[d-1]}$ projects us onto a unique vacuum, and the resulting theory should have symmetry
\be
\hat{K}_{[-1]}\times G_{[0]}
\ee
where $G_{[0]}$ is non-anomalous.  The original anomaly has been shifted away from $G_{[0]}$ -- just as in the 1d example above, we reproduce it by putting $G_{[0]}$ on the background defect labeled by $\hat{k}$, where it acquires the anomaly $\chi_{\hat{k}}(c(g_1,g_2,...,g_{d+1}))=\omega(g_1,...,g_{d+1})$.

\subsection{2-Group Resolution in 2d}
\label{sec:2group}

As an additional example of the above procedure, we can return to two dimensions.  Previously, in section~\ref{sec:z22d}, we resolved an anomalous $(\Z_2)_{[0]}$ by extending it to $(\Z_4)_{[0]}$.  Now instead we consider the extended symmetry
\be
\label{2gext}
(\Z_2)_{[1]}.(\Z_2)_{[0]}
\ee
with extension class $c$ being the non-trivial element in $H^3(\Z_2,\Z_2)=\Z_2$.  Such an extension of a 0-form symmetry by a 1-form symmetry is known as a 2-group, and the extension class is called the Postnikov class \cite{Baez}.  The Postnikov, valued in $H^3(G,K)$ can be understood in a manner similar to the $H^3(G,U(1))$-valued 't Hooft anomalies we have been examining so far: in swapping a four-way junction of TDLs labeled by $G$ (which, being constrained to obey the group law, depends on only three elements of $G$) we pick up a local operator labeled by an element of $K$, rather than simply a phase.  The Postnikov tells us exactly which local operator appears in the swap relation for each triple of $G$ elements.

When gauging a 2-group symmetry, then, we should expect each sector of the orbifold torus partition function to be decorated not just by a line wrapping each homotopy cycle, but additionally by the possible presence of a local operator $\sigma_k$.  Thus we will use the notation $Z_{g_1,g_2;k}$ to denote such a configuration, as shown in Figure~\ref{fig:2gpf}.

\begin{figure}
	\centering
	\begin{tikzpicture}
	\draw[thin] (0,0)--(5,0);
	\draw[thin] (5,0)--(5,5);
	\draw[thin] (0,0)--(0,5);
	\draw[thin] (0,5)--(5,5);
	\draw[very thick,->] (1.5,0)--(2,1.25);
	\draw[very thick] (2,1.25)--(2.5,2.5);
	\draw[very thick,->] (2.5,2.5)--(3,3.75);
	\draw[very thick] (3,3.75)--(3.5,5);
	\draw[very thick] (4.25,2.5)--(5,2.5);
	\draw[very thick,->] (2.5,2.5)--(4.25,2.5);
	\draw[very thick] (2.5,2.5)--(1.5,2.5);
	\draw[very thick] (0.75,2.5)--(1.5,2.5);
	\draw[very thick,->] (0,2.5)--(0.75,2.5);
	\node at (2.5,4) {$g_1$};
	\node at (2.5,1) {$g_1$};
	\node at (1,2.8) {$g_2$};
	\node at (4.5,2.8) {$g_2$};
	\filldraw[black] (3.5,1.5) circle (2pt);
	\node at (3.75,1.25) {$\sigma_k$};
	\end{tikzpicture}
	\caption{The visual representation of $Z_{g_1,g_2;k}$.}
	\label{fig:2gpf}
\end{figure}

In the example at hand with 2-group symmetry $(\Z_2)_{[1]}.(\Z_2)_{[0]}$, the orbifold partition function would come as a sum over eight sectors:
\be
\label{2gpf}
\frac{1}{4}[Z_{0,0;0}+Z_{0,0;1}+Z_{0,1;0}+Z_{0,1;1}+Z_{1,0;0}+Z_{1,0;1}+Z_{1,1;0}+Z_{1,1;1}]
\ee
with $\Z_2$ written additively.

Let us take a moment to appreciate how the answer would differ if instead of (\ref{2gext}) we had a direct product symmetry, i.e.~if the Postnikov class were trivial.  We would then have (due to the 1-form symmetry) a direct sum of two copies of a theory with an anomalous $(\Z_2)_{[0]}$ symmetry.  Of course, due to the anomaly, we would be unable to consistently gauge the $\Z_2$ in either or both copies.  The presence of the non-trivial extension class in (\ref{2gext}) allows us to absorb the anomalous phases that would have appeared when swapping $\Z_2$ TDLs into the definition of the local operators that appear due to the 1-form symmetry.  This means that, for instance, the final six terms of (\ref{2gpf}), which in the direct product would come as two modular invariant pairs of three terms each, now forms a single modular orbit, as they obey relations such as
\be
Z_{1,1;0}(\tau+1)=Z_{1,0;1}(\tau).
\ee

Note that this procedure is in line with the calculations of \cite[section 3.2]{BeniniCordovaHsin} which shows that the 't Hooft anomaly of a 0-form symmetry in 2d is given by a quotient of $H^3(G,U(1))$ by $H^3(G,K)\cup \hat{K}$.  That is, the presence of a non-trivial 2-group extension has the ability to trivialize part of the 0-form anomaly.  As explained above, the interpretation of the $K$ character appearing here is that it labels a background defect associated to the $-1$-form symmetry that would appear when gauging $K_{[1]}$.

\subsection{Resolving an Anomalous 1-form Symmetry in 3d}

The procedures that we've examined so far have dealt exclusively with resolving anomalies in 0-form symmetries.  For our final example we will instead examine anomalies in 1-form symmetries in 3d theories and attempt to engineer a method of resolving them.\footnote{For additional perspectives on resolving anomalies in higher-form symmetries, see \cite{WanWang}.}

1-form symmetries in 3d are described by topological line operators.  Their 't Hooft anomalies are given by phases which appear when we swap a pair of crossing lines \cite{BeniniCordovaHsin}.  In particular, letting the 1-form symmetry be given by the group $G$, the process of straightening out a loop in a line may pick up a phase $q(g)$, referred to as the topological spin of the $g$ line.  $q$ is a quadratic function mapping $G$ to $U(1)$.  From $q$ we can get the phase appearing in the crossing of any two lines $g_1$ and $g_2$ as
\be
\frac{q(g_1g_2)}{q(g_1)q(g_2)}.
\ee

If any lines have non-trivial topological spin, the 1-form symmetry will be anomalous in the sense that it will be impossible to couple the theory to a background network of $G$ lines in a gauge-invariant manner \cite{Hsin_2019}.  The possible maps $q$ are classified by $H^4(B^2G,U(1))$ \cite{KapustinThorngren}, where $B^2G$ is the second Eilenberg-MacLane space for $G$, i.e.~a topological space $X$ with $\pi_2(X)=G$ and all other homotopy trivial.

In order to construct a symmetry extension that resolves such an anomaly, note that extensions of a 1-form symmetry $G_{[1]}$ by a 2-form symmetry $K_{[2]}$ are classified by $H^4(B^2G,K)$ \cite{Tachikawa}.  Letting $c\in H^4(B^2G,K)$ be the extension class, then, we choose an abelian group $K$ and a character $\hat{k}\in\hat{K}$ such that
\be
c\cup \hat{k} = q
\ee
with $q$ regarded as an element of $H^4(B^2G,U(1))$.  We then expect to be able to resolve the anomaly $q$ in $G_{[1]}$ via extending to a total symmetry given by
\be
\label{k2g1}
K_{[2]}.G_{[1]}
\ee
with extension class $c$.

In the theory with symmetry (\ref{k2g1}), straightening out a loop in a line produces a local operator, quite similarly to the case (examined in the previous subsection) of a 2-group in 2d, where swap relations among lines produced local operators.  The anomaly can then be absorbed as a phase redefinition of these local operators.  Gauging $K_{[2]}$ leads to a theory with symmetry
\be
\hat{K}_{[-1]}\times G_{[1]}
\ee
with non-anomalous 1-form symmetry $G$ and non-trivial mixed anomaly.  The remainder of the explanation proceeds identically to the 0-form case; in particular, the mixed anomaly reproduces the original 't Hooft anomaly on the background defect labeled by $\hat{k}$.

\newpage

\section{Conclusion}

In summary, the procedure outlined in section~\ref{sec:2d} extends a 0-form symmetry $G$ by an abelian 0-form symmetry $K$.  Gauging $K$ produces a theory which has a non-anomalous $G$ symmetry, which is possible because this $G$ now has an action on the defects added by the introduction of $K$.  The original anomaly now manifests in a mix of $G$ and the quantum $(d-2)$-form symmetry $\hat{K}$.  In section~\ref{sec:1d}, $K$ was instead a $(d-1)$-form symmetry.  Once again the $K$-gauged theory has a non-anomalous $G$ symmetry.  The original $G$ anomaly manifests when the theory is put on a particular background defect $\hat{k}$ arising from the $-1$-form symmetry dual to $K$.  We also saw that a similar procedure can resolve 1-form anomalies in three dimensions.\\

The concrete examples we saw were, for tractability and simplicity, mostly focused on lower dimensions.  Given that, as we increase dimension, there is an increasing variety of types of defects which can exist and carry various types of 't Hooft anomalies, it would be interesting to investigate additional resolution schema beyond those analyzed here.  (As a quick example, one might ask for a method of resolving a mixed anomaly between mixed-form symmetries.)  Following the interpretation presented in section~\ref{sec:tqft}, a related pursuit would be an investigation of the various types of TQFTs one could couple to a given theory, and what the degrees of freedom in those couplings are.

\appendix

\section{Fusion Categorical Extensions}
\label{appendix1}

We would like to examine extensions of the form
\be
1 \lrr \Rep(G) \lrr \Rep(\Gamma) \lrr \Rep(K) \lrr 1
\ee
where the constituents are the fusion categories (see \cite{ENO} for a mathematical exposition and \cite{BhardwajTachikawa,Chang:2018iay} for physics-oriented overviews) formed by fusion of the irreducible representations of the groups $G$, $K$ and $\Gamma$.  The map $\io:\Rep(G)\to\Rep(\Gamma)$ is inclusion, which (much as in the group-like case) can be specified by an injective map from the simple objects of $\Rep(G)$ to the simple objects of $\Rep(\Gamma)$.  We can then take $\pi:\Rep(\Gamma)\to\Rep(K)$ to be a linear map on the objects of $\Rep(\Gamma)$ for which simple objects of $\Rep(\Gamma)$ are taken to either the identity in $\Rep(K)$ or to a not necessarily simple object in $\Rep(K)$ which does not include the identity \cite{BruguieresNatale}.  In particular, $\pi(\io(\hat{g}))=d_{\hat{g}}1_{\Rep(K)}$ for $\hat{g}$ a simple object of dimension $d_{\hat{g}}$ in $\Rep(G)$.  $\pi$ should additionally respect the fusion rules, in that 
\be
\pi(\hat{\gamma}_1\otimes_{\Rep{\Gamma}}\hat{\gamma}_2)=\pi(\hat{\gamma}_1)\otimes_{\Rep(K)}\pi(\hat{\gamma}_2)
\ee
for $\hat{\gamma}_1,\hat{\gamma}_2\in\Rep(\Gamma)$.  Finally, all simple objects of $\Rep(K)$ should appear in the image of at least one simple object of $\Rep(\Gamma)$.

For the simplest example, note that $S_3$ fits into the short exact sequence
\be
1 \lrr \Z_3 \lrr S_3 \lrr \Z_2 \lrr 1,
\ee
which means we expect a dual short exact sequence
\be
\label{reps3_es}
1 \lrr \Z_2 \lrr \Rep(S_3) \lrr \Z_3 \lrr 1.
\ee
$\Rep(S_3)$ has three simple objects: 1, $X$ and $Y$ of dimension 1, 1 and 2 with fusion rules
\begin{center}
\begin{tabular}{c || c | c | c}
$\otimes$ & 1 & $X$ & $Y$ \\\hline\hline
1 & 1 & $X$ & $Y$ \\\hline
$X$ & $X$ & 1 & $Y$ \\\hline
$Y$ & $Y$ & $Y$ & $1\oplus X\oplus Y$ \\
\end{tabular}.
\end{center}
Clearly 1 and $X$ generate a $\Ve_{\Z_2}$ (that is, $\Z_2$ viewed as a fusion category) subsymmetry.  Writing $\Z_3$ additively as $\{0,1,2\}$, we can take $\pi(1)=\pi(X)=0$, $\pi(Y)=1\oplus 2$.  There is a single non-trivial calculation required to check that such a map respects the fusion rules.  We compare
\be
\pi(Y\otimes Y)=\pi(1\oplus X\oplus Y)=\pi(1)\oplus\pi(X)\oplus\pi(Y)=0\oplus 0\oplus 1\oplus 2
\ee
to
\be
\pi(Y)\otimes\pi(Y) = (1\oplus 2)\otimes(1\oplus 2) = 0 \oplus 0 \oplus 1 \oplus 2.
\ee
Thus, we have constructed the maps appearing in the short exact sequence (\ref{reps3_es}).\\

As a more involved example, we can consider $S_4$, which has two non-trivial normal subgroups: $\Z_2\times\Z_2$ and $A_4$.  Considering first the short exact sequence
\be
1 \lrr A_4 \lrr S_4 \lrr \Z_2 \lrr 1,
\ee
we expect $\Rep(S_4)$ to fit into
\be
1 \lrr \Z_2 \lrr \Rep(S_4) \lrr \Rep(A_4) \lrr 1.
\ee
To construct the associated maps, we will need the fusion rules.  $\Rep(S_4)$ has five simple objects: 1, $S$, $P$, $Q$ and $R$ of dimension 1, 1, 2, 3 and 3 with fusion rules
\begin{center}
\begin{tabular}{c || c | c | c | c | c}
$\otimes$ & 1 & $S$ & $P$ & $Q$ & $R$ \\\hline\hline
1 & 1 & $S$ & $P$ & $Q$ & $R$ \\\hline
$S$ & $S$ & 1 & $P$ & $R$ & $Q$ \\\hline
$P$ & $P$ & $P$ & $1\oplus S\oplus P$ & $Q\oplus R$ & $Q\oplus R$ \\\hline
$Q$ & $Q$ & $Q$ & $Q\oplus R$ & $1\oplus P\oplus Q\oplus R$ & $S\oplus P\oplus Q\oplus R$ \\\hline
$R$ & $R$ & $R$ & $Q\oplus R$ & $S\oplus P\oplus Q\oplus R$ & $1\oplus P\oplus Q\oplus R$
\end{tabular},
\end{center}
and $\Rep(A_4)$ has four simple objects: $1$, $A$, $B$, $C$ of dimension 1, 1, 1 and 3 with fusion rules
\begin{center}
\begin{tabular}{c || c | c | c | c}
$\otimes$ & 1 & $A$ & $B$ & $C$ \\\hline\hline
1 & 1 & $A$ & $B$ & $C$ \\\hline
$A$ & $A$ & $B$ & 1 & $C$ \\\hline
$B$ & $B$ & 1 & $A$ & $C$ \\\hline
$C$ & $C$ & $C$ & $C$ & $1\oplus A\oplus B\oplus 2C$
\end{tabular}.
\end{center}
Once again, the inclusion map is obvious as $\{1,S\}\in\Rep(S_4)$ form a $\Z_2$ subsymmetry.  For the map $\pi$, we can take $\pi(1)=\pi(S)=1$, $\pi(P)=A\oplus B$ and $\pi(Q)=\pi(R)=C$.  A few of the less trivial checks on this choice of map would be
\be
\pi(Q\otimes P) = \pi(Q\oplus R) = \pi(Q) \oplus \pi(R) = 2C
\ee
versus
\be
\pi(Q)\otimes\pi(P) = C\otimes(A\oplus B) = 2C
\ee
and 
\be
\pi(Q\otimes R) = \pi(S\oplus P \oplus Q \oplus R) = 1 \oplus A\oplus B \oplus 2C
\ee
versus
\be
\pi(Q)\otimes\pi(R) = C\otimes C = 1\oplus A \oplus B \oplus 2C.
\ee\\

$S_4$ also fits into the short exact sequence
\be
1 \lrr \Z_2\times\Z_2 \lrr S_4 \lrr S_3 \lrr 1,
\ee
so we would expect to find
\be
1 \lrr \Rep(S_3) \lrr \Rep(S_4) \lrr \Z_2\times\Z_2 \lrr 1.
\ee
Examining the $\Rep(S_4)$ fusion rules, it is clear that $\{1,S,P\}$ generate a $\Rep(S_3)$ subsymmetry, so again it remains to identify an appropriate map $\pi$.  Writing $\Z_2\times\Z_2=\{1,a,b,c\}$, let us take $\pi(1)=\pi(S)=1$, $\pi(P)=2$ and $\pi(Q)=\pi(R)=a\oplus b\oplus c$.  Some consistency checks on this choice are given by comparing
\be
\pi(Q\otimes Q)=\pi(1\oplus P\oplus Q\oplus R) = 3\oplus 2a\oplus 2b\oplus 2c
\ee
with
\be
\pi(Q)\otimes\pi(Q) = (a\oplus b\oplus c)\otimes(a\oplus b\oplus c) = 3\oplus 2a\oplus 2b\oplus 2c
\ee
and
\be
\pi(Q\otimes P) = \pi(Q\oplus R) = \pi(Q)\oplus \pi(R) = (a\oplus b\oplus c)\oplus (a\oplus b\oplus c)
\ee
with
\be
\pi(Q)\otimes\pi(P) = (a\oplus b\oplus c) \otimes 2 = 2a\oplus 2b\oplus 2c.
\ee

Interestingly, there is another consistent set of fusion rules closely related to $\Rep(S_4)$, where we have objects of the same dimension but fusion rules \cite{Liu_2022}
\begin{center}
\begin{tabular}{c || c | c | c | c | c}
$\otimes$ & 1 & $S$ & $P$ & $Q$ & $R$ \\\hline\hline
1 & 1 & $S$ & $P$ & $Q$ & $R$ \\\hline
$S$ & $S$ & 1 & $P$ & $R$ & $Q$ \\\hline
$P$ & $P$ & $P$ & $1\oplus S\oplus P$ & $Q\oplus R$ & $Q\oplus R$ \\\hline
$Q$ & $Q$ & $Q$ & $Q\oplus R$ & $S\oplus P\oplus Q\oplus R$ & $1\oplus P\oplus Q\oplus R$ \\\hline
$R$ & $R$ & $R$ & $Q\oplus R$ & $1\oplus P\oplus Q\oplus R$ & $S\oplus P\oplus Q\oplus R$
\end{tabular},
\end{center}
where the difference between this category's fusion rules and those of $\Rep(S_4)$ is the bottom-right 2x2 block, that is the fusion of $Q$ and $R$ among themselves.  Note that in $\Rep(S_4)$ we have $Q^*=Q$ and $R^*=R$, but the fusion rules above would lead to $Q^*=R$ and $R^*=Q$, where the asterisk denotes dualization (the dual $X^*$ of a simple object $X$ is the unique simple object for which $X \otimes X^*$ contains the identity; in a group-like fusion category this would be the inverse).

This alternative fusion category (which, it should be noted, is not $\Rep(G)$ for any group $G$) would still be expressible as either an extension of $\Z_2\times\Z_2$ by $\Rep(S_3)$ or an extension of $\Rep(A_4)$ by $\Z_2$, using the maps above.  This would seem to stem from the freedom that we have in exchanging the roles of $1$ and $S$, both simple objects of dimension 1 appearing in the kernel of $\pi$.  Note that in the previous examples the simple objects in the kernel of $\pi$ appeared symmetrically, such that exchanging them would have had no effect on the fusion rules.

\addcontentsline{toc}{section}{References}

\bibliographystyle{utphys}
\bibliography{AnomResolution}

\providecommand{\href}[2]{#2}\begingroup\raggedright\begin{thebibliography}{10}

\bibitem{WWW}
J.~Wang, X.-G. Wen, and E.~Witten, ``Symmetric gapped interfaces of {SPT} and
  {SET} states: Systematic constructions,''
  \href{http://dx.doi.org/10.1103/physrevx.8.031048}{{\em Physical Review X}
  {\bfseries 8} no.~3, (Aug, 2018) }.
  \url{https://doi.org/10.1103%2Fphysrevx.8.031048}.

\bibitem{TopOps}
D.~Robbins, E.~Sharpe, and T.~Vandermeulen, ``Decomposition, trivially-acting
  symmetries, and topological operators,'' 2022.
\newblock \url{https://arxiv.org/abs/2211.14332}.

\bibitem{lfs}
T.~Vandermeulen, ``Lower-form symmetries,'' 2022.
\newblock \url{https://arxiv.org/abs/2211.04461}.

\bibitem{Kobayashi_2019}
R.~Kobayashi, K.~Ohmori, and Y.~Tachikawa, ``On gapped boundaries for {SPT}
  phases beyond group cohomology,''
  \href{http://dx.doi.org/10.1007/jhep11(2019)131}{{\em Journal of High Energy
  Physics} {\bfseries 2019} no.~11, (Nov, 2019) }.
  \url{https://doi.org/10.1007%2Fjhep11%282019%29131}.

\bibitem{KapustinSeiberg}
A.~Kapustin and N.~Seiberg, ``Coupling a {QFT} to a {TQFT} and duality,''
  \href{http://dx.doi.org/10.1007/jhep04(2014)001}{{\em Journal of High Energy
  Physics} {\bfseries 2014} no.~4, (Apr, 2014) }.
  \url{https://doi.org/10.1007%2Fjhep04%282014%29001}.

\bibitem{Seiberg_2010}
N.~Seiberg, ``Modifying the sum over topological sectors and constraints on
  supergravity,'' \href{http://dx.doi.org/10.1007/jhep07(2010)070}{{\em Journal
  of High Energy Physics} {\bfseries 2010} no.~7, (Jul, 2010) }.
  \url{https://doi.org/10.1007%2Fjhep07%282010%29070}.

\bibitem{GaiottoKulp}
D.~Gaiotto and J.~Kulp, ``Orbifold groupoids,''
  \href{http://dx.doi.org/10.1007/jhep02(2021)132}{{\em Journal of High Energy
  Physics} {\bfseries 2021} no.~2, (Feb, 2021) }.
  \url{https://doi.org/10.1007%2Fjhep02%282021%29132}.

\bibitem{Unsal}
M.~Ünsal, ``Strongly coupled {QFT} dynamics via {TQFT} coupling,''
  \href{http://dx.doi.org/10.1007/jhep11(2021)134}{{\em Journal of High Energy
  Physics} {\bfseries 2021} no.~11, (Nov, 2021) }.
  \url{https://doi.org/10.1007%2Fjhep11%282021%29134}.

\bibitem{SP1}
T.~Pantev and E.~Sharpe, ``Notes on gauging noneffective group actions,'' 2005.
\newblock \url{https://arxiv.org/abs/hep-th/0502027}.

\bibitem{SP2}
T.~Pantev and E.~Sharpe, ``String compactifications on calabi{\textendash}yau
  stacks,'' \href{http://dx.doi.org/10.1016/j.nuclphysb.2005.10.035}{{\em
  Nuclear Physics B} {\bfseries 733} no.~3, (Jan, 2006) 233--296}.
  \url{https://doi.org/10.1016%2Fj.nuclphysb.2005.10.035}.

\bibitem{SP3}
T.~Pantev and E.~Sharpe, ``G{LSM}s for gerbes (and other toric stacks),''
  \href{http://dx.doi.org/10.4310/atmp.2006.v10.n1.a4}{{\em Advances in
  Theoretical and Mathematical Physics} {\bfseries 10} no.~1, (2006) 77--121}.
  \url{https://doi.org/10.4310%2Fatmp.2006.v10.n1.a4}.

\bibitem{VafaDT}
C.~Vafa and E.~Witten, ``On orbifolds with discrete torsion,''
  \href{http://dx.doi.org/10.1016/0393-0440(94)00048-9}{{\em Journal of
  Geometry and Physics} {\bfseries 15} no.~3, (Feb, 1995) 189--214}.
  \url{https://doi.org/10.1016%2F0393-0440%2894%2900048-9}.

\bibitem{ChenGuLiuWen}
X.~Chen, Z.-C. Gu, Z.-X. Liu, and X.-G. Wen, ``Symmetry protected topological
  orders and the group cohomology of their symmetry group,''
  \href{http://dx.doi.org/10.1103/physrevb.87.155114}{{\em Physical Review B}
  {\bfseries 87} no.~15, (Apr, 2013) }.
  \url{https://doi.org/10.1103%2Fphysrevb.87.155114}.

\bibitem{HsinLam}
P.-S. Hsin and H.~T. Lam, ``Discrete theta angles, symmetries and anomalies,''
  \href{http://dx.doi.org/10.21468/scipostphys.10.2.032}{{\em {SciPost}
  Physics} {\bfseries 10} no.~2, (Feb, 2021) }.
  \url{https://doi.org/10.21468%2Fscipostphys.10.2.032}.

\bibitem{LOZ}
L.~Li, M.~Oshikawa, and Y.~Zheng, ``Non-invertible duality transformation
  between spt and ssb phases,'' 2023.
\newblock \url{https://arxiv.org/abs/2301.07899}.

\bibitem{DW}
R.~Dijkgraaf and E.~Witten, ``{Topological gauge theories and group
  cohomology},'' \href{http://dx.doi.org/cmp/1104180750}{{\em Communications in
  Mathematical Physics} {\bfseries 129} no.~2, (1990) 393 -- 429}.
  \url{https://doi.org/}.

\bibitem{UndoingDecomp}
E.~Sharpe, ``Undoing decomposition,''
  \href{http://dx.doi.org/10.1142/s0217751x19502336}{{\em International Journal
  of Modern Physics A} {\bfseries 34} no.~35, (Dec, 2019) 1950233}.
  \url{https://doi.org/10.1142%2Fs0217751x19502336}.

\bibitem{FRS}
J.~Fuchs, I.~Runkel, and C.~Schweigert, ``{TFT construction of RCFT correlators
  1. Partition functions},''
  \href{http://dx.doi.org/10.1016/S0550-3213(02)00744-7}{{\em Nucl. Phys. B}
  {\bfseries 646} (2002) 353--497},
  \href{http://arxiv.org/abs/hep-th/0204148}{{\ttfamily arXiv:hep-th/0204148}}.

\bibitem{BhardwajTachikawa}
L.~Bhardwaj and Y.~Tachikawa, ``{On finite symmetries and their gauging in two
  dimensions},'' \href{http://dx.doi.org/10.1007/JHEP03(2018)189}{{\em JHEP}
  {\bfseries 03} (2018) 189}, \href{http://arxiv.org/abs/1704.02330}{{\ttfamily
  arXiv:1704.02330 [hep-th]}}.

\bibitem{Chang:2018iay}
C.-M. Chang, Y.-H. Lin, S.-H. Shao, Y.~Wang, and X.~Yin, ``Topological defect
  lines and renormalization group flows in two dimensions,''
  \href{http://dx.doi.org/10.1007/JHEP01(2019)026}{{\em JHEP} {\bfseries 01}
  (2019) 026}, \href{http://arxiv.org/abs/1802.04445}{{\ttfamily
  arXiv:1802.04445 [hep-th]}}.

\bibitem{Tachikawa}
Y.~Tachikawa, ``On gauging finite subgroups,''
  \href{http://dx.doi.org/10.21468/scipostphys.8.1.015}{{\em {SciPost} Physics}
  {\bfseries 8} no.~1, (Jan, 2020) }.
  \url{https://doi.org/10.21468%2Fscipostphys.8.1.015}.

\bibitem{BruguieresNatale}
A.~Bruguières and S.~Natale, ``Exact sequences of tensor categories,'' 2010.
\newblock \url{https://arxiv.org/abs/1006.0569}.

\bibitem{ENO_extensions}
P.~Etingof, D.~Nikshych, V.~Ostrik, and w.~a. a. b.~E. Meir, ``Fusion
  categories and homotopy theory,'' 2009.
\newblock \url{https://arxiv.org/abs/0909.3140}.

\bibitem{AEO}
D.~G. Robbins, E.~Sharpe, and T.~Vandermeulen, ``Anomalies, extensions, and
  orbifolds,'' \href{http://dx.doi.org/10.1103/physrevd.104.085009}{{\em
  Physical Review D} {\bfseries 104} no.~8, (Oct, 2021) }.
  \url{https://doi.org/10.1103%2Fphysrevd.104.085009}.

\bibitem{DecompIntro}
E.~Sharpe, ``An introduction to decomposition,'' 2022.
\newblock \url{https://arxiv.org/abs/2204.09117}.

\bibitem{Wang_2018}
J.~Wang, K.~Ohmori, P.~Putrov, Y.~Zheng, Z.~Wan, M.~Guo, H.~Lin, P.~Gao, and
  S.-T. Yau, ``Tunneling topological vacua via extended operators:
  (spin-){TQFT} spectra and boundary deconfinement in various dimensions,''
  \href{http://dx.doi.org/10.1093/ptep/pty051}{{\em Progress of Theoretical and
  Experimental Physics} {\bfseries 2018} no.~5, (May, 2018) }.
  \url{https://doi.org/10.1093%2Fptep%2Fpty051}.

\bibitem{Baez}
J.~C. Baez and A.~D. Lauda, ``Higher-dimensional algebra v: 2-groups,''.
  \url{https://arxiv.org/abs/math/0307200}.

\bibitem{BeniniCordovaHsin}
F.~Benini, C.~C\'ordova, and P.-S. Hsin, ``On 2-group global symmetries and
  their anomalies,'' \href{http://dx.doi.org/10.1007/JHEP03(2019)118}{{\em
  JHEP} {\bfseries 03} (2019) 118},
  \href{http://arxiv.org/abs/1803.09336}{{\ttfamily arXiv:1803.09336
  [hep-th]}}.

\bibitem{WanWang}
Z.~Wan and J.~Wang, ``{Adjoint QCD$_4$, Deconfined Critical Phenomena,
  Symmetry-Enriched Topological Quantum Field Theory, and Higher
  Symmetry-Extension},''
  \href{http://dx.doi.org/10.1103/PhysRevD.99.065013}{{\em Phys. Rev. D}
  {\bfseries 99} no.~6, (2019) 065013},
  \href{http://arxiv.org/abs/1812.11955}{{\ttfamily arXiv:1812.11955
  [hep-th]}}.

\bibitem{Hsin_2019}
P.-S. Hsin, H.~T. Lam, and N.~Seiberg, ``Comments on one-form global symmetries
  and their gauging in 3d and 4d,''
  \href{http://dx.doi.org/10.21468/scipostphys.6.3.039}{{\em {SciPost} Physics}
  {\bfseries 6} no.~3, (Mar, 2019) }.
  \url{https://doi.org/10.21468%2Fscipostphys.6.3.039}.

\bibitem{KapustinThorngren}
A.~Kapustin and R.~Thorngren, ``Higher symmetry and gapped phases of gauge
  theories,'' 2013.
\newblock \url{https://arxiv.org/abs/1309.4721}.

\bibitem{ENO}
P.~Etingof, D.~Nikshych, and V.~Ostrik, ``On fusion categories,'' 2002.
\newblock \url{https://arxiv.org/abs/math/0203060}.

\bibitem{Liu_2022}
Z.~Liu, S.~Palcoux, and Y.~Ren, ``Classification of grothendieck rings of
  complex fusion categories of multiplicity one up to rank six,''
  \href{http://dx.doi.org/10.1007/s11005-022-01542-1}{{\em Letters in
  Mathematical Physics} {\bfseries 112} no.~3, (Jun, 2022) }.
  \url{https://doi.org/10.1007%2Fs11005-022-01542-1}.

\end{thebibliography}\endgroup

\end{document}